\def\papertitle{ A Diffusion-based generative equalizer for music restoration}

\def\paperauthorA{Eloi Moliner}
\def\paperauthorB{Maija Turunen}
\def\paperauthorC{Filip Elvander}
\def\paperauthorD{Vesa Välimäki}

\documentclass[twoside,a4paper]{article}
\usepackage{etoolbox}
\usepackage{dafx_24}
\usepackage{amsmath,amssymb,amsfonts,amsthm, bbm}
\usepackage{euscript}
\usepackage[T1]{fontenc}
\usepackage[utf8]{inputenc}
\usepackage{ifpdf}
\usepackage[english]{babel}
\usepackage{caption}
\usepackage{subfig} 
\usepackage{algorithm}
\usepackage{algpseudocode}
\usepackage{bm}
\usepackage{color}
\usepackage{booktabs}

\input glyphtounicode
\ninept

\newcounter{numauth}
\newcounter{listcnt}
\newcommand\authcnt[1]{\ifdefined#1 \stepcounter{numauth} \fi}

\newcommand\addauth[1]{
\ifdefined#1 
\stepcounter{listcnt}
\ifnum \value{listcnt}<\value{numauth}
\appto\authorslist{, #1}
\else
\appto\authorslist{~and~#1}
\fi
\fi}
\authcnt{\paperauthorB}
\authcnt{\paperauthorC}
\authcnt{\paperauthorD}
\authcnt{\paperauthorE}
\authcnt{\paperauthorF}
\authcnt{\paperauthorG}
\authcnt{\paperauthorH}
\authcnt{\paperauthorI}
\authcnt{\paperauthorJ}
\def\authorslist{\paperauthorA}
\addauth{\paperauthorB}
\addauth{\paperauthorC}
\addauth{\paperauthorD}
\addauth{\paperauthorE}
\addauth{\paperauthorF}
\addauth{\paperauthorG}
\addauth{\paperauthorH}
\addauth{\paperauthorI}
\addauth{\paperauthorJ}

\usepackage{times}

\newif\ifpdf
\ifx\pdfoutput\relax
\else
   \ifcase\pdfoutput
      \pdffalse
   \else
      \pdftrue
\fi

\ifpdf 
  \usepackage[pdftex,
    pdftitle={\papertitle},
    pdfauthor={\authorslist},
    pdfsubject={Proceedings of the 27th International Conference on Digital Audio Effects (DAFx24)},
    colorlinks=false, 
    bookmarksnumbered, 
    pdfstartview=XYZ 
  ]{hyperref}
  \usepackage[pdftex]{graphicx}
\else 
  \usepackage[dvips]{epsfig,graphicx}
  \usepackage[dvips,
    pdftitle={\papertitle},
    pdfauthor={\authorslist},
    pdfsubject={Proceedings of the 27th International Conference on Digital Audio Effects (DAFx24)},
    colorlinks=false, 
    bookmarksnumbered, 
    pdfstartview=XYZ 
  ]{hyperref}
\fi
\usepackage[hypcap=true]{caption}
\title{\papertitle}

\usepackage{svg}
\usepackage{varwidth}
\usepackage{pgf}
\usepackage{pgfplots}
\usepackage{pgfplotstable}
\pgfplotsset{compat=1.3}
\usepackage[skins]{tcolorbox}
\usepackage{adjustbox}

\affiliation{
\paperauthorA $^{1}$,
\paperauthorB $^{2}$,
\paperauthorC $^{1}$ and 
\paperauthorD $^{1}$\thanks{\vspace{-3mm}}\hspace{-2pt}
}
{
$^1$\href{https://www.aalto.fi/en/aalto-acoustics-lab}{Acoustics Lab}, Department of Information and Communications Engineering, Aalto University, Espoo, Finland \\
$^2${\href{https://www.uniarts.fi/en/units/sibelius-academy/}{Sibelius Academy}, University of the Arts Helsinki, Helsinki, Finland  \\
}
{\tt \href{mailto:eloi.moliner@aalto.fi}{eloi.moliner@aalto.fi}}
}

\begin{document}
\ifpdf 
  \DeclareGraphicsExtensions{.png,.jpg,.pdf}
\else  
  \DeclareGraphicsExtensions{.eps}
\fi


\maketitle

\begin{abstract}
This paper presents a novel approach to audio restoration, focusing on the enhancement of low-quality music recordings, and in particular historical ones. 
Building upon a previous algorithm called BABE, or Blind Audio Bandwidth Extension, we introduce BABE-2, which presents a series of improvements.
This research broadens the concept of bandwidth extension to \emph{generative equalization}, a novel task that, to the best of our knowledge, has not been explicitly addressed in previous studies. 
BABE-2 is built around an optimization algorithm utilizing priors from diffusion models, which are trained or fine-tuned using a curated set of high-quality music tracks. The algorithm simultaneously performs two critical tasks: estimation of the filter degradation magnitude response and hallucination of the restored audio. The proposed method is objectively evaluated on historical piano recordings, showing an enhancement over the prior version. The method yields similarly impressive results in rejuvenating the works of renowned vocalists Enrico Caruso and Nellie Melba. 
This research represents an advancement in the practical restoration of historical music.
Historical music restoration examples are available at: 
\href{http://research.spa.aalto.fi/publications/papers/dafx-babe2/}{research.spa.aalto.fi/publications/papers/dafx-babe2/}.


\end{abstract}

\section{Introduction}
\vspace{-10pt}
\label{sec:intro}

Historical music recordings suffer from severe impairment due to limitations of the physical recording media, alongside the wear and tear from playback and storage \cite{godsill_digital_1998, kob2019interprete}.
These degradations can take the form of additive disturbances, decreased bandwidth, coloration, and nonlinear effects \cite{Valimaki2008}.
This project goal is to enhance the quality of 
historical music recordings  the audio standards of today, making them more accessible for people used to modern sound quality.
 Restoring these recordings is challenging for several reasons. First, the problem is ill-posed, 
, with a multitude of potential solutions, thereby complicating the quest for the optimal one.
 Additionally, the degradation process is unknown, making the restoration a blind inverse problem for which the solution must be found without definite information on the necessary repairs.




In the 1970s, Stockham et al.~began groundbreaking efforts in digitally restoring gramophone recordings, notably those of the celebrated singer Enrico Caruso \cite{stockham1975blind}.
At the time, the approach was limited by technological constraints, relying on the estimation of an equalization curve derived from the spectral average of a contemporary recording. 
Since then, the field of audio restoration has evolved significantly \cite{godsill_digital_1998}.
Today, data-driven techniques employing deep neural networks can remove all additive disturbances simultaneously \cite{moliner2022two}. Deep generative models, such as diffusion models \cite{lemercier2024diffusion}, address complex restoration tasks, such as bandwidth extension \cite{moliner2023zeroshot}, with unprecedented success. This study investigates a critical question:
``\emph{Can deep generative models elevate the quality of historical music recordings to modern standards?}''


To aid in this investigation, we introduce and explore the concept of \emph{generative equalization}, a task designed to overcome the limitations of traditional equalization on ill-conditioned scenarios, i.e., when some frequency bands are not present in the original recording.
 In such cases, the equalization curve must be cropped; otherwise, it will only amplify background noise.
 In contrast, a generative equalizer can synthesize the missing spectral components as to achieve the target spectral profile, as it is a generalization of bandwidth extension.
 While the combination of equalization and bandwidth extension has been previously studied \cite{Qian2004combining}, this paper, to the best of our knowledge, is the first to explore this approach in the context of generative models and music restoration.
 
This work builds upon BABE, the authors' prior research in audio bandwidth extension \cite{moliner2023zeroshot}, advancing it within the framework of generative equalization. 
While this method showed promise for addressing lowpass degradation, its application to real-world historical recordings revealed significant limitations, primarily due to an oversimplified degradation model. 
The model inadequately described the complex coloration effects often present in historical recordings, effects that are typically attributed to the recording technology, such as the recording horn \cite{Valimaki2008, kob2019interprete}.

This paper introduces BABE-2, an improvement and extension of the original BABE model, bringing a set of technical contributions designed to refine its efficacy for the restoration of historical music recordings. This includes incorporating a novel parametrization of the degradation filter model as a zero-phase frequency response estimate, allowing for a more flexible and richer filter structure extending beyond the previously constrained lowpass characteristic. Additionally, we implement a regularization strategy to counteract the breakpoint-collapse problem encountered in BABE, introduce noise regularization to foster stable convergence, and propose an initialization scheme based on the long-term average spectra (LTAS), alongside other critical implementation details.



This paper is structured as follows.
Sec.~\ref{sec:background} outlines the basic principles required 
to follow the description of the new algorithm. This section describes the basics of diffusion models, diffusion posterior sampling, and the previous BABE algorithm.  
Following this, Sec.~\ref{sec:babe2} delves into the enhancements made in the BABE-2 method, expanding the bandwidth extension techniques introduced in \cite{moliner2023zeroshot} to encompass the broader concept of generative equalization.
Sec.~\ref{sec:results} reports on experiments conducted with historical recordings of piano and singing voice,
and Sec.~\ref{sec:voice_analysis} describes our procedure for selecting the training data used for restoring specific historical singers.
Furthermore, this section elucidates on some of the capabilities and limitations of BABE-2 for singing voice restoration.
Finally, Sec.~\ref{sec:conclusions} concludes the paper.

%
%
%





\section{Background}\label{sec:background}


\subsection{Diffusion Models}
Diffusion models have emerged as a powerful class of generative models, demonstrating remarkable capabilities in various domains, including image and audio processing \cite{song2020score, karras2022elucidating,lemercier2024diffusion}. 
These models operate by gradually transforming data through a process of adding and removing noise. In the context of audio, we define time as the variable \(\tau\), with \(\mathbf{x}_\tau\) representing the state of the audio data at time \(\tau\). At \(\tau = T\), \(\mathbf{x}_T\) is distributed as Gaussian noise, while \(\mathbf{x}_0\) represents the clean audio signal.
Following the formulation by Karras et al.~\cite{karras2022elucidating}, the underlying Ordinary Differential Equation 
governing this transformation is written as
\begin{equation}\label{odekarras}
    d\mathbf{x}_\tau=  - \tau  \nabla_{\mathbf{x}_\tau}\log p_\tau(\mathbf{x}_\tau) d\tau.
\end{equation}
The \emph{score} function $\nabla_{\mathbf{x}_\tau}\log p_\tau(\mathbf{x}_\tau) d\tau$, 
 which guides the model to reverse the noise addition process, is approximated using a denoiser $D_\theta(\mathbf{x}_\tau, \tau)$, a deep neural network with parameters $\theta$:
\begin{equation} \label{score}
    \nabla_{\mathbf{x}_\tau}\log p_\tau(\mathbf{x}_\tau) \approx (D_\theta(\mathbf{x}_\tau,\tau)-\mathbf{x}_\tau)/\sigma(\tau)^2,
\end{equation}
where $\sigma(\tau)^2$ is the noise variance at timestep $\tau$\footnote{Following Karras et al. \cite{karras2022elucidating}, we employ $\sigma(\tau)=\tau$. }.
For training these models, Denoising Score Matching is often employed, which aims to minimize the difference between the denoised and the original clean audio. This objective is encapsulated as follows:
 \begin{equation}\label{loss}
    \mathbb{E}_{\mathbf{x}_0 \sim p_\text{data}, \boldsymbol\epsilon \sim \mathcal{N}(\mathbf{0},\mathbf{I}) }  \left[ \lambda(\tau) \lVert D_\theta(\mathbf{x}_0+\tau\mathbf{\epsilon},\tau) -\mathbf{x}_0   \rVert_2^2 \right], 
\end{equation}
where $\lambda(\tau)$ is a weighting parameter defined according to \cite{karras2020analyzing}.
In the subsequent sections,
the term $\hat{\mathbf{x}}_0$ denotes the denoised estimate obtained through the application of neural network $D_\theta$.

\subsection{Diffusion Posterior Sampling}

Many restoration tasks can be understood as inverse problems, where an observed measurement signal is obtained by applying a certain degradation operator
$H(\cdot)$ 
to an unknown signal $\mathbf{x}_0$, yielding the observation
$\mathbf{y}=H(\mathbf{x}_0)$.
Diffusion models are highly effective as data-driven priors for solving inverse problems, where the goal is to estimate the original signal $\mathbf{x}_0$ from the degraded observations $\mathbf{y}$  \cite{moliner2022solving}.
When the degradation operator
$H(\cdot)$  
is known, these models can be used to approximate the posterior distribution 
$p(\mathbf{x}_0|\mathbf{y})$. Ideally, to solve an inverse problem using diffusion models, one can modify the ODE in Eq.~\eqref{odekarras}, replacing the score with $\nabla_{\mathbf{x}_\tau}\log p_\tau(\mathbf{x}_\tau|\mathbf{y})$, also known as the \emph{posterior score}. 

We adopt the ``Diffusion Posterior Sampling'' method introduced by Chung et al. \cite{chung2022diffusion}, which utilizes a Bayesian framework, decomposing the posterior score into two parts:
 \begin{equation} \label{posterior}
\nabla_{\mathbf{x}_\tau}\log p_\tau(\mathbf{x}_\tau|\mathbf{y})=
 \nabla_{\mathbf{x}_\tau}\log p_\tau(\mathbf{x}_\tau)+
     \nabla_{\mathbf{x}_\tau}\log p_\tau(\mathbf{y}|\mathbf{x}_\tau).
\end{equation}
The first term, the \emph{prior score}, is derived from Eq.~\eqref{score}, while the second, the \emph{likelihood score}, is estimated through the gradients of an application-specific cost function:
\begin{equation}\label{recguid}
    \nabla_{\mathbf{x}_\tau}\log p_\tau(\mathbf{y}|\mathbf{x}_\tau) \approx
    -\xi(\tau) \; \nabla_{\mathbf{x}_\tau} 
 C_\text{audio}(\mathbf{y}, H(\hat{\mathbf{x}}_0))
 ,
\end{equation}
where $\xi(\tau)$ serves as a gradient normalization factor \cite{moliner2022solving}, and $C_\text{audio}(\cdot, \cdot)$ denotes an appropriately chosen cost function.

Within this framework, $\hat{\mathbf{x}}_0=D_\theta(\mathbf{x}_\tau,\tau)$ represents the Minimum Mean Squared Error (MMSE) estimate of the clean audio $\mathbf{x}_0$ given the noisy or transformed state $\mathbf{x}_\tau$.
It is important to note that implementing this method necessitates differentiating through the denoiser, which is a deep neural network.
The differentiation step adds a layer of computational complexity to the process. 
{This} approach has been successfully used for solving audio restoration tasks, such as bandwidth extension, inpainting, and declipping \cite{moliner2022solving, hernandez2024vrdmg}.

\subsection{Application to Blind Inverse Problems}

In scenarios where the degradation operator $H$ is unknown, the challenges in solving the inverse problem increase significantly. These are referred to as blind inverse problems.
Some recent works offer solutions for solving blind inverse problems with diffusion models by jointly optimizing the degradation operator alongside the data during the sampling process \cite{chung2022parallel}. 
The methodology relies on alternating between updates from diffusion sampling and optimization steps for finding the unknown operator.


Our recent work BABE \cite{moliner2023zeroshot}, focusing on the blind bandwidth extension of music signals without knowledge of the lowpass filter characteristics, fits in this framework.
In this approach, the degradation operator $H_\phi$ is modeled as a zero-phase filter in the frequency domain, with a piecewise linear magnitude response defined by a set of parameters $\phi$. The parameters are iteratively optimized through stochastic gradient descent while sampling from the diffusion model in a coarse-to-fine manner.

In BABE, the filter parameters are constrained to ensure they describe a strictly decaying function \cite{moliner2023zeroshot}. 
The reason for doing that was to limit the filter design to a lowpass characteristic, aiming to stabilize the optimization convergence.
Such a parameterization, however, does not accommodate spectral coloration or resonances, which are common in historical recordings and can significantly impact the audio quality. We hypothesize that the constrained modeling capability degrades the performance of posterior sampling in real historical recordings due to unaddressed spectral properties, pushing the samples towards out-of-distribution regions compared to the training data and thus reducing the reliability of diffusion model predictions.





\section{BABE-2: Unique Contributions} \label{sec:babe2}

\subsection{Filter Parameterization} \label{sec:filter_parameterization}


In addressing blind inverse problems within a posterior sampling framework, it is essential to specify a set of pertinent degradation operators. Following our earlier approach \cite{moliner2023zeroshot}, the focus is on zero-phase frequency-domain forward filters denoted as $H_\phi$.
The relationship between the observations $\mathbf{y}$ and the reconstructed estimate $\hat{\mathbf{x}}_0$ through these filters is given by:
\begin{equation}
   \mathbf{y} \approx \mathcal{F}^{-1}( H_\phi \odot \mathcal{F}(\hat{\mathbf{x}}_0)), 
\end{equation}
where $\mathcal{F}(\cdot)$ and $\mathcal{F}^{-1}(\cdot)$ represent the Fourier transform and its inverse, respectively. We use the FFT length of 4096 samples. For simplicity, we use the notation $H_\phi(\hat{\mathbf{x}}_0)$ interchangeably to describe this operation throughout the paper.



\begin{figure}
     \centering
    \includegraphics{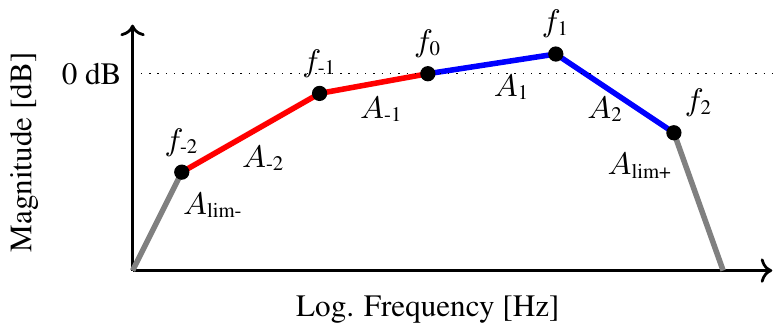}
    \vspace{-5pt}
    \caption{\it{Proposed frequency-response equalizer model consists of breakpoints creating a piecewise linear magnitude response.}}
    \vspace{-10pt}
    
    \label{fig:babe}
\end{figure}

The lowpass filter operator, initially applied in the BABE framework \cite{moliner2023zeroshot}, is extended to a parametric frequency-response equalizer. The equalizer configuration, depicted in Fig.~\ref{fig:babe}, incorporates an adjustable anchor cutoff frequency ($f_0$) with a fixed magnitude at 0\,dB to ensure stability in the filter's gain across its operational bandwidth. 
The design includes a set of adjustable breakpoints and slopes ($f_i$ and $A_i$) localized above and below the anchor frequency, similarly to those employed in BABE but with the flexibility of non-strictly decaying slopes. 
The first and last breakpoints in in Fig.~\ref{fig:babe} delineate the effective spectral range, assuming there is no significant spectral energy beyond these limits.
The slopes before and after the limiting breakpoints are not optimizable and are fixed to $A_\text{lim-}=80$\,dB and $A_\text{lim+}=-80$\,dB, respectively.
We refer to Appendix A.4
\footnote{Appendix available in the pre-print: \href{https://arxiv.org/abs/2403.18636}{arxiv.org/abs/2403.18636} }
for further details on the filter paramaterization.
The filter is piecewise differentiable with respect to the cutoff frequencies and slope parameters.

Choosing the right number of filter stages is crucial, especially under the assumption that the magnitude response should exhibit smoothness. 
Employing a filter with more stages provides greater flexibility during optimization but risks overfitting to minor spectral details. This overfitting can lead to an unrealistic magnitude response that could negatively impact the optimization.
With this in mind,
a smaller number of stages is favored to maintain smoothness, and we have chosen a configuration with 5 breakpoints, as depicted in Fig~\ref{fig:babe}.

To mitigate potential training instabilities, we impose restrictions on the slopes, ensuring that they do not exceed predefined minimum and maximum values of -40 dB and 40 dB, respectively. 
Additionally, the cutoff frequencies are confined within the range $(f_\text{min}, f_\text{max})$, which is determined by the FFT size employed in the filter parameterization. 
In our experiments, we set $f_\text{min}=10$\;Hz and $f_\text{max}=\frac{1}{2}f_\text{s}$ to the Nyquist limit, depending on the sampling rate. 
Furthermore, we enforce that a sequential ordering of the cutoff frequencies,  requiring each frequency to be less than the subsequent one.
These constraints are implemented through the application of projections within the specified ranges during the optimization process. Ideally, the constraint on the ordering of cutoff frequencies would be unnecessary, as the regularization scheme introduced in the next subsection aims to prevent parameter configurations that could lead to such scenarios.

\subsection{Breakpoint-Collapse Regularization}\label{sec:bcr}


We identify a previously unrecognized problem in the BABE algorithm \cite{moliner2023zeroshot}, which we term \emph{breakpoint collapse}.
The problem occurs when, during inference, two or more cutoff frequency breakpoints converge to very close values, receiving nearly identical gradients.
 Consequently, the breakpoints cannot separate from each other for the rest of the process, reducing the flexibility of the degradation model by losing these breakpoint and slope parameters.


Interestingly, this problem was not prevalent in simulated lowpass filtered recordings as tested in BABE. However, it became significantly more frequent and problematic when the model was applied to real historical recordings, particularly those that differed much from the training data distribution. 
To address this problem, we introduce a novel regularization term, the Breakpoint-Collapse Regularization (BCR). The BCR term is expressed as
\begin{equation}
    \mathcal{C}_\text{BCR}(\phi)\!=\!
    e^{-\beta (f_{\text{min}}-f_{\text{-}S^\prime})}+
    \!\!\sum_{i=\text{-}S^\prime}^{S-1}\!\! e^{-\beta (f_{i}-f_{(i+1)})}+
    e^{-\beta (f_{S}- f_{\text{max}})}.
\end{equation}


\noindent This function imposes a minimal cost when breakpoints are adequately spaced but enforces an exponentially increasing cost, when they approach each other. The parameter $\beta$, set to 0.1 in our implementation, modulates this behavior. 

To integrate this into our framework, the total cost function is modified to include the BCR term, weighted appropriately. The revised total cost equation is
\begin{equation}
\mathcal{C_\text{total}}=\mathcal{C_\text{filter}}(\mathbf{y},H_\phi(\hat{\mathbf{x}}_0))+\gamma_\text{BCR}\mathcal{C}_\text{BCR}(\phi),
\end{equation}
where $\gamma_\text{BCR}$, the  BCR term weight, is set to 10 in our experiments.

\subsection{Noise Regularization} \label{sec:noise_reg}

Our parametric model, designed to fit linear magnitude responses, faces challenges in addressing nonlinear artifacts within historical audio recordings. 
These artifacts, often made worse by preprocessing steps like denoising, can bias the optimization of magnitude response parameters, leading to convergence at local minima and suboptimal restoration outcomes.
This problem was overlooked in the previous iteration of the method \cite{moliner2023zeroshot}. 
As we show in Sec.~\ref{sec:piano_eval}, the performance of BABE on historical recordings is suboptimal.

To solve these problems, we suggest adding a certain amount of stochasticity, or ``noise,'' into the optimization objective. We do this by adding random noise to our observations, which we describe as  $\bar{\mathbf{y}}=\mathbf{y}+\gamma \bm{\epsilon}$, where $\gamma$ is the noise scale and $\bm{\epsilon} \sim \mathcal{N}(\mathbf{0}, \mathbf{I})$ is a vector of Gaussian noise.
Technically, this approach smooths the likelihood distribution by convolving it with a Gaussian kernel. This trick enables the optimization process to navigate the cost function landscape more effectively than without it.
Each iteration's unique noise ensures the optimizer does not prematurely converge to local minima, thereby encouraging a thorough search for optimal solutions. 
Furthermore, the additive noise potentially masks some of the artifacts that may be present in the observed data, preventing these artifacts from negatively affecting the optimization. This approach aligns with techniques used by Sadat et al.~\cite{sadat2023cads}, which aimed to enhance diversity in image generation.  
In Sec.~\ref{sec:piano_eval}, we conduct an ablation study where this technique is not applied, which is there denoted as ``BABE-2 w/o noise reg.''

\subsection{Long-term Average Spectrum-based Initialization}\label{sec:LTAS}

A known blind equalization technique designs a filter to align the input LTAS with that of a reference set \cite{stockham1975blind}.
To achieve this, we calculate the reference LTAS, denoted as $\tilde{\mathbf{R}}_\text{LTAS}$, by applying the Short-Time Fourier Transform (STFT) to the reference set and average the power spectra across frames. In the STFT, we use 4096-sample windows at the sampling rate of $f_s=44.1$\,kHz, or 2048-sample windows at $f_s=22.05$\,kHz. We employ a hop size of 1/4 in both cases.
One-third-octave band Gaussian smoothing is applied to the averaged spectra. 
The LTAS of the input $\tilde{\mathbf{Y}}_\text{LTAS}$ is determined in the same way, based on the original recording.

By computing the ratio between the LTAS spectra, the equalization filter is obtained as
\begin{equation}\label{eq:ratio}
    |H_\text{LTAS}|=\tilde{|\mathbf{Y}}_\text{LTAS}|/|\tilde{\mathbf{R}}_\text{LTAS}|,
\end{equation}
where $H_\text{LTAS}$ is an estimate of the forward degradation. 
The inverse filter $H_\text{LTAS}^{-1}$ can be used as a frequency-domain equalizer:
\begin{equation} \label{eq:LTASeq}
    \tilde{\mathbf{y}}= 
    \mathcal{F}^{-1}(
    H_\text{LTAS}^{-1} \odot \mathcal{F}(\mathbf{y})
    ),
\end{equation}
where $\odot$ is the Hadamard product.

From now on, this equalization method is referred to as ``LTAS-EQ.'' While this approach proved useful for correcting significant colorations \cite{stockham1975blind}, it has strong limitations.
Firstly, if the observations are bandlimited, the ratio in Eq. \eqref{eq:ratio} is ill-conditioned, as it approaches zero for some frequency bands, causing the inverse filter to amplify to extremely high values. 
To mitigate this, we limit $H_\text{LTAS}$ at $-20$\,dB. Secondly, an estimate involving time averages assumes the measured signal is ergodic, which may often not be true. The LTAS method overlooks the dynamics of the audio, such as the unique sound of a piano when played at different intensities. Consequently, during LTAS computation, louder sections may disproportionately influence the average spectrum, leading to an imbalance. Due to these limitations, relying solely on this approach for audio restoration is not recommended.


Considering these limitations, we adopt this method as a preliminary step that BABE-2 will refine. Our goal is to facilitate the equalization process, aiming for a more stable optimization of the filter at the initial stages and achieving convergence with fewer iterations. We explore two approaches to accomplish this.
Firstly, we utilize $H_\text{LTAS}$ for an improved initial setup of the audio signal at starting time $T$, $\mathbf{x}_T$. Unlike in BABE \cite{moliner2023zeroshot}, where $\mathbf{x}_T$ began as a noisy version of the observations $\mathbf{x}_T \sim \mathcal{N}({\mathbf{y}},\sigma_\text{start}^2\mathbf{I})$, we now introduce a modified approach. We suggest beginning with a noisy version of the equalized observations $\mathbf{x}_T \sim \mathcal{N}({\tilde{\mathbf{y}}},\sigma_\text{start}^2\mathbf{I})$, where $\tilde{\mathbf{y}}$ is derived using Eq.~\eqref{eq:LTASeq}. This version of our method is hereon referred to as ``BABE-2 w/ LTAS-EQ init.''

The second strategy extends the use of LTAS-EQ beyond initial setup by incorporating it into the optimization objective. This involves replacing the original recording $\mathbf{y}$ with $\tilde{\mathbf{y}}$ in the reconstruction cost functions from Eqs.~\eqref{eq:rec} and \eqref{eq:filter_rec}. It is crucial to recognize that in this context, the estimated frequency response $H_\phi$ no longer corresponds directly to the original recording response $\mathbf{y}$. Instead, the connection between the original recording and the estimated restored version $\hat{\mathbf{x}}_0$ becomes a composite of $H\phi$ and $H_\text{LTAS}$ as formalized below:
\begin{equation}
   \mathbf{y} \approx \mathcal{F}^{-1}
   \left( 
   (H_\phi \odot H_\text{LTAS} )
   \odot    \mathcal{F}(\hat{\mathbf{x}}_0) 
   \right).
\end{equation}
 We denote this variant as ``BABE-2 w/ LTAS-EQ obj.''

\subsection{Improved Inference Algorithm}

\begin{algorithm}[t]
\caption{Inference phase of the BABE-2 method 
}
\label{alg3}
\begin{algorithmic}
\Require observations $\mathbf{y}$
\State $\mathbf{x}_{T}\leftarrow$ Warm Initialization$(\mathbf{y})$ \Comment{Initialize audio signal}
\State Initialize $\phi_{T}$ \Comment{initialize the filter parameters}
\For{$i \leftarrow T, \dots, 1$} \Comment{discrete step backwards}
\State $\hat{\mathbf{x}}_0 \leftarrow D_\theta(\mathbf{x}_i ,\sigma_i) $ \Comment{evaluate denoiser}

\State Optimize $\phi_i$ \Comment{Optimization loop}
\State Likelihood score approximation ($\approx \nabla_{\mathbf{x}_i} \log p(\mathbf{y}| \mathbf{x}_i, \phi_i)$)
\State Update $ \mathbf{x}_{i}$
\EndFor
\State \Return $\mathbf{x}_0$  \Comment{reconstructed audio signal}
\end{algorithmic}
\end{algorithm}

The inference algorithm of BABE-2 shares strong similarities with the one employed in BABE \cite{moliner2023zeroshot}. 
The pseudocode in Algorithm \ref{sec:appendix_details} summarizes the inference process, highlighting the new contributions in blue. 
Despite, for brevity reasons, Algorithm 1 shows an adaptation of a $1^\text{st}$ order Euler sampler, we use the $2^\text{nd}$ order stochastic sampler introduced in \cite{karras2022elucidating}.
One relevant addition is the use of the Adam optimizer \cite{kingma_adam_2017} to optimize the filter parameters, instead of a simpler stochastic gradient descent as used in \cite{moliner2023zeroshot}. Adam applies adaptive per-parameter learning rates and second-order moments of the gradients, which stabilizes the performance for non-convex optimization problems with sparse gradients. 

As the reconstruction cost function for the audio signal, we employ the same standard $L_2$ norm as used in BABE:
\begin{equation} \label{eq:rec}
    C_\text{audio}(\mathbf{y}, H_\phi(\hat{\mathbf{x}}_0))=
    \lVert
    \mathbf{y} -
    H_\phi(\mathbf{\hat{x}}_0)
    \rVert_2^2.
\end{equation}
In contrast, we now employ a slightly different reconstruction cost function for optimizing the filter parameters. 
BABE-2 uses a high-frequency emphasized $L_2$ norm, denoted as
\begin{equation}\label{eq:filter_rec}
  C_\text{filter}(\mathbf{y}, H_\phi(\hat{\mathbf{x}}_0) )=
  \lVert 
  H_\text{emph.}(\mathbf{y}-H_\phi
  (\hat{\mathbf{x}}_0))
  \rVert_2^2,
\end{equation}
where $H_\text{emph.}$ is a first-order high-pass pre-emphasis filter \cite{wright2020perceptual}.
Further implementation details are omitted from the main paper, but they are given in the Appendix A and in the source code \footnote{\href{https://github.com/eloimoliner/BABE2}{github.com/eloimoliner/BABE2}}.

\section{Experiments and Evaluation} \label{sec:results}

We experiment with the restoration of solo piano and singing recording. The conducted objective evaluation compares the performance of BABE-2 against comprehensive baselines for enhancing the quality of historical 78-RPM gramophone recordings.

\subsection{Piano Recordings Evaluation}\label{sec:piano_eval}
 \begin{figure*}[t]
     \centering
     \includegraphics[width=\textwidth]{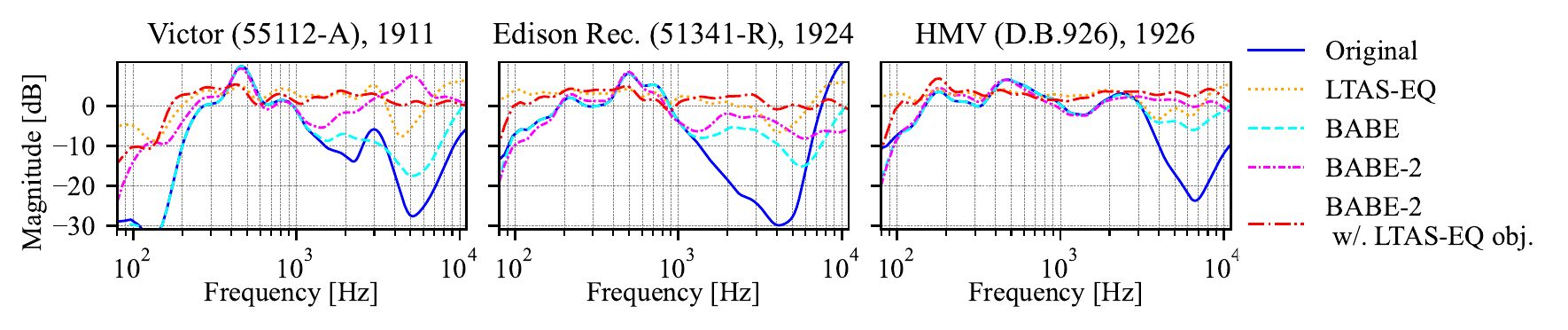}
     \vspace{-10pt}
     \caption{\textit{Comparative LTAS analysis of original and restored piano recordings using different methods.}}
     \vspace{-5pt}
        
     \label{fig:ltas}
 \end{figure*}

Here we assess the effectiveness of our proposed method for restoring historical piano recordings.
For ease of comparison, we used the same diffusion model as the one used in BABE \cite{moliner2023zeroshot}. This model is based on the CQT-Diff+ architecture \cite{moliner2022diffusion}, and was trained on the MAESTRO dataset \cite{hawthorne2018enabling}, resampled at 22\;kHz.  

The test set comprises 54 gramophone recordings of solo piano performances from 1910 to 1930, all of them collected from ``the Internet Archive''\footnote{\href{https://archive.org/details/78rpm}{archive.org/details/78rpm}}. Each recording was trimmed to a duration of 1 min. These recordings were preprocessed using our denoiser \cite{moliner2022two} to eliminate additive noise disturbances, such as clicks, hisses, and pops. The processed evaluation set is made available as part of the supplementary materials\footnote{\href{http://research.spa.aalto.fi/publications/papers/dafx-babe2/}{research.spa.aalto.fi/publications/papers/dafx-babe2/}}.

The objective evaluation is based on the Fréchet Audio Distance (FAD), a reference-free metric comparing the statistics between two sets of embeddings, as detailed by Kilgour et al. \cite{kilgour2019frechet}. The efficacy of FAD is influenced by the specific embeddings employed and the chosen reference set \cite{gui2023adapting}. To provide a thorough analysis, we computed FAD across various embeddings utilizing the `fadtk` library \cite{gui2023adapting}. We began with the VGGish classifier' embeddings, widely used in this context \cite{kilgour2019frechet}, and additionally assessed the model using CLAP, a joint audio–text representation, \cite{wu2023large}, and Encodec, designed for low-rate audio codecs \cite{defossez2022high}.
As a reference test set, we used the MAESTRO dataset test and validation splits from the year 2018.

Furthermore, we examined the distance between the LTAS of the restored recordings and the LTAS of the reference test set. We acknowledge that a distance based on LTAS is not a definitive metric for the same reasons as outlined in Sec.~\ref{sec:LTAS}. However, we include it here as it can shed light on significant differences in coloration with respect to the reference set.
The LTAS distance is measured as follows:
\begin{equation}
\text{LTAS dist.}=10\log_{10} 
   \left(
   \frac{1}{K}
   \sum_f\frac{
   |
   \tilde{\mathbf{X}}_\text{LTAS}-
   \tilde{\mathbf{R}}_\text{LTAS}
   |
   }
   {
   \tilde{\mathbf{R}}_\text{LTAS}
   } 
   \right),
\end{equation}
where \(K\) denotes the number of frequency bins, and \(\tilde{\mathbf{X}}_\text{LTAS}\) and \(\tilde{\mathbf{R}}_\text{LTAS}\) represent the LTAS of the restored recordings and the reference test set, respectively.

The results are presented at the top of Table 1. The compared versions of BABE-2 consistently outperform the original recordings, the LTAS-EQ and BEHM-GAN baselines \cite{moliner2022behm}, and the predecessor BABE \cite{moliner2023zeroshot}.
An ablation study highlights the critical role of noise regularization for better performance. Additionally, the results show that employing LTAS-based initialization strategies improves the FAD scores. 


To delve deeper into the impact of the restoration methods on different recordings, Fig.~\ref{fig:ltas} illustrates the frequency-response estimates derived from LTAS analysis.
This figure includes three plots corresponding to distinct solo piano recordings, each labeled with its respective year and identifier, depicting the magnitude in dB across a logarithmically spaced frequency axis.
The blue lines represent the frequency-response estimates for the original recordings, calculated based on the ratio between the LTAS of each recording and that of the reference test set, following Eq.~\eqref{eq:ratio}.
The remaining lines in each plot denote the LTAS ratio for each restoration method, as detailed in the legend. 

Ideally, in Fig.~\ref{fig:ltas}, an equalized recording should exhibit a relatively flat frequency response, centered around 0 dB. 
It can be observed that the original recordings' responses possess an irregular shape, characterized by deficiencies in both high and low-frequency energy and marked coloration, i.e.~fluctuations, in the mid-frequency range. The ``LTAS-EQ'' method manages to mitigate the coloration to some extent but, in doing so, it amplifies background noise and distortion artifacts.
This harms the FAD scores, as evidenced in Table 1.
The ``BABE'' method, indicated by the cyan dashed line, increases the high-frequency energy, yet not enough to align with the target equalization curve. In contrast, the ``BABE-2'' method achieves a more balanced equalization, particularly effective when combined with the LTAS-EQ objective. 
 This analysis showcases the superiority of the BABE-2 method.
 



\begin{table}[t]
\caption{\textit{Objective evaluation on historical recordings. The best results of each column for each experiment are bolded.}}
\vspace{-5pt}
\label{tab:results}
\resizebox{\columnwidth}{!}{%
\begin{tabular}{@{}l|lll|l@{}}
\toprule
                       & \multicolumn{3}{c|}{FAD $\downarrow$} & LTAS  \\ 
Experiment, method        & VGGish & CLAP & Encodec  &  dist. $\downarrow$          \\ \midrule
\textit{Piano}: Original              & 2.37   & 0.33 & 8.11     & -1.15 dB   \\
LTAS-EQ                  & 3.33   & 0.19 & 8.12     & -1.77 dB   \\
BEHM-GAN              & 1.82          & 0.21 & 6.65         & -1.37 dB \\
BABE                  & 1.50   & 0.15 & 7.72     & -2.68 dB   \\
BABE-2                & 1.45     & \textbf{0.12}   & 4.65   & -3.02 dB   \\
BABE-2 w/o noise reg.    & 1.46   & 0.15 & 7.25     & -2.37 dB   \\
BABE-2 w/ LTAS-EQ init.     & 1.50          & \textbf{0.12} & \textbf{4.56}      & \textbf{-3.42 dB}   \\
BABE-2 w/ LTAS-EQ obj.     & \textbf{1.20}       & \textbf{0.12} & 5.21      & -2.57 dB   \\ \midrule
\textit{Vocals (Caruso)}: Original           & 19.34     & 0.47 & 31.01    & -0.29 dB   \\
LTAS-EQ               & 19.59     & 0.52 & 23.41    & \textbf{-1.96 dB}   \\
BABE               & 14.48     & 0.31 & 26.10    & -1.06 dB   \\
BABE-2              & 11.36     & 0.28 & 21.08    & -1.65 dB   \\
BABE-2 w/ LTAS-EQ init.   & 11.32            & 0.29 & 20.53    & -1.84 dB   \\
BABE-2 w/ LTAS-EQ obj.     & \textbf{11.11}       & \textbf{0.27} & \textbf{15.04}    & -1.95 dB   \\ \midrule
\textit{Vocals (Melba)}: Original           & 11.56     & 0.63 &  41.34    & -0.94 dB   \\
LTAS-EQ               & 11.99     & 0.66 & 37.04    & \textbf{-2.39 dB}   \\
BABE               & 5.33     & 0.36 &  32.68   & -0.88 dB   \\
BABE-2              & 3.71     & 0.31 & 27.78    & -0.84 dB   \\
BABE-2 w/ LTAS-EQ init.   & \textbf{3.44}            & \textbf{0.30} & \textbf{22.12}    & -1.56  dB   \\
BABE-2 w/ LTAS-EQ obj.     & 3.58       & 0.32 & 23.64    & -2.23 dB   \\ \bottomrule
\end{tabular}%
}

\end{table}



 \begin{figure*}[t]
     \centering
    \includegraphics{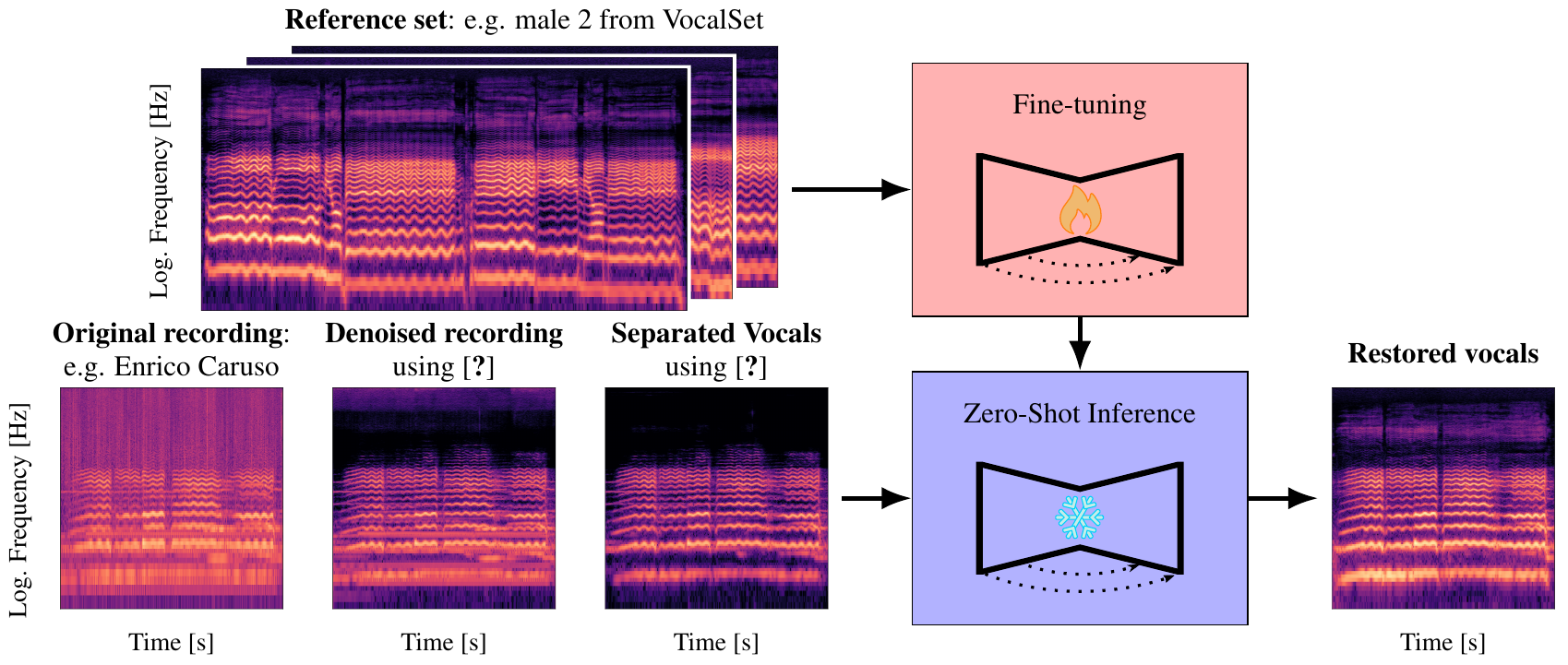}
    \vspace{-5pt}    \caption{\textit{Singing voice restoration pipeline.}}
    \vspace{-10pt}
    
    \label{fig:sv_pipeline}
\end{figure*}

\subsection{Evaluation of Singing Voice Recordings} \label{sec:singing_eval}


We evaluate our restoration method to recordings from two iconic singers: Enrico Caruso (tenor) and Nellie Melba (soprano).
For this purpose, we pretrained a diffusion model with a collection of studio quality modern singing voice recordings from different sources\footnote{See Appendix A for more information regarding the training data.}, all of them sampled at 44.1 kHz.
This model was later fine-tuned for each target singer in particular, using smaller reference sets extracted from selected singers from Vocalset \cite{wilkins2018vocalset}, 
as we elaborate in Sec.~\ref{sec:voice_analysis}.
We opted for standard fine-tuning, resuming the training from the pretrained weights.



The process for restoring singing voice recordings is shown in Fig.~\ref{fig:sv_pipeline}. Similar to the approach detailed in Sec.~\ref{sec:piano_eval}, all recordings were trimmed to 1 min and denoised  \cite{moliner2022two} as an initial step. Our evaluation is exclusively aimed at enhancing vocal tracks, and hence, the instrumental parts were separated using HT-Demucs \cite{rouard2023hybrid}. While this is effective, the source separation method is not fully optimized for historical recordings and tends to introduce noticeable artifacts. These artifacts, however, are outside the scope of this study, as we are not focusing on source separation.

We collected two evaluation sets consisting of 32 recordings from Enrico Caruso and 25 from Nellie Melba.
The objective evaluation mirrors the one conducted for piano music, employing the same metrics for assessment. The outcomes for both Caruso and Melba are detailed in Table \ref{tab:results}. For the Caruso recordings, the BABE-2 version utilizing the LTAS-EQ objective achieved the most favorable FAD scores. On the other hand, for Melba's recordings, the BABE-2 version with LTAS-EQ initialization showed slightly superior performance.
It is noteworthy that, despite recording significantly lower FAD scores, the LTAS-EQ baseline registered the smallest LTAS distance.

\section{Discussion: Restoring historical voices}\label{sec:voice_analysis}
\begin{figure*}[t]
    \centering
    \includegraphics{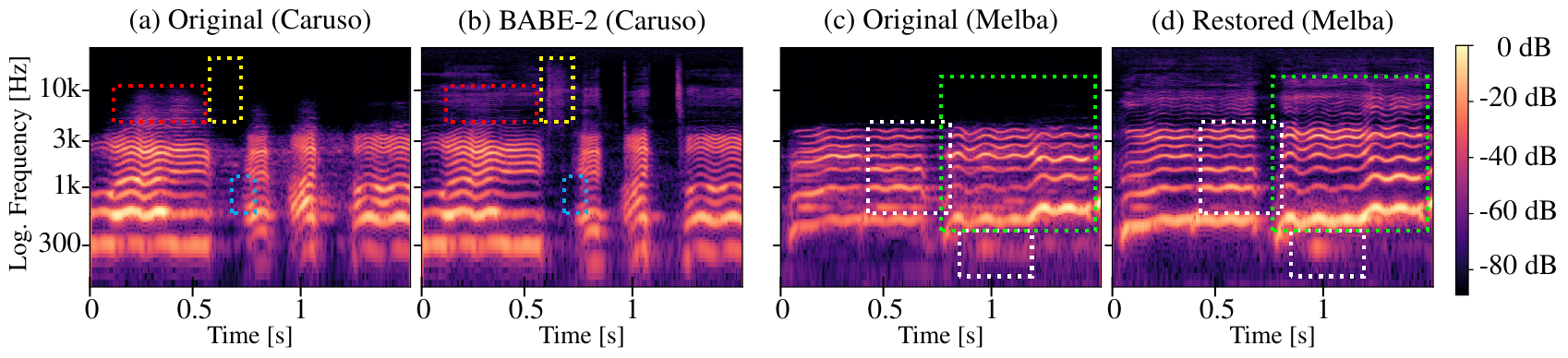}
    \vspace{-5pt}
    \caption{\textit{Spectrogram representations of two vocal restoration examples. The colored boxes highlight key points discussed in Sec.~\ref{sec:qualitativeanalysis}.}}
    \vspace{-5pt}
    \label{fig:spectrograms}
\end{figure*}

One of the key contributions from our research lies in our efforts to restore early-20th-century recordings of classical opera singers.
These recordings are often so much degraded that some of the unique vocal characteristics of the singer are no longer preserved.
The role of BABE-2 is to reconstruct, or hallucinate, the missing pieces, based on statistical patterns derived from the training data.
As a result, the final output can be considered a hybrid containing features from both the original recording and the training data. 
This underscores the critical importance of selecting appropriate training data, as it significantly impacts the accuracy and authenticity of the audio restoration, ensuring the reconstructed voices closely mirror the original timbres and nuances.

Our methodology involves the strategic fine-tuning of diffusion models, which have been pre-trained on a broad spectrum of singing voice recordings, with a more focused dataset of high-quality recordings from a unique and carefully selected reference singer. The reference singers are chosen for their vocal qualities, which closely resemble those of the historical figures we seek to restore. For this purpose, we employ VocalSet \cite{wilkins2018vocalset}, a dataset comprising studio-quality a cappella recordings from 20 professional singers. Each singer contributed with approximately 30 min of data, encompassing a diverse array of vocal techniques.

\subsection{Searching for Reference Voices}

The task of identifying voices that resemble a particular opera singer is not trivial. Opera singers can be divided into different voice categories, “fach”, based on, for example, vocal range and timbre, permeability, or flexibility of their voices \cite{Steane2002}.
To get the full impression of the singer's voice characteristics on a historical recording, it is necessary not only to listen but also to consult the writings of contemporaries and to study the singer's repertoire. 
Enrico Caruso (1871-1921) was known for his distinctive voice and his interpretations of tenor roles in the Italian opera repertoire for example Cavaradossi from Puccini’s \textit{Tosca} or Canio from \textit{Pagliacci} by Leoncavallo \cite{freestone1961enrico}.
Both roles require stamina and permeability in low and high register, and Caruso's voice has been described as having a baritonal quality in the low and brilliance of the tenor voice in the higher register. Caruso's recording career spanned 20 years and during that time his technique stabilized, and his voice matured and gained depth \cite{gentili2021birth, Celletti2013}.
Therefore, we only seek an average estimate of his voice.

We found that the voice from VocalSet's \textit{Male-2} possesses a similar warmth and depth in the low and middle registers as Caruso, together with the power and clarity in the higher register, hence we used it for fine-tuning Caruso's recordings.
The famous aria “Vesti la giubba” from Pagliacci,  included among the audio examples on our webpage, is a perfect example of Caruso’s voice but also of his dramatic abilities. 
On the webpage, you also find a version of the same aria sung by Beniamino Gigli (1890-1957), who was considered a legitimate heir of Caruso and their voices have similarities, but Gigli's voice sounds more metallic \cite{Shawe-Taylor2001Gigli}.
Gigli's restored example was produced using the dramatic voice from VocalSets's \textit{Male-11}.

Nellie Melba (1861–1931) was complimented on the beauty of her voice and her flawless technique with equalized scale from top to bottom. Her colleague John McCormack described her voice as a lyric soprano with a beautiful tone and her technique as perfection. She had the ability to sing coloraturas with bird-like quality, and her voice was light and agile \cite{strong1941john}.
The beauty of voice is a subjective matter, but we found a similar light lyric quality in the voice from VocalSet's \textit{Female-1}, which we used for fine-tuning. 
Like Caruso, Melba had a long recording career, and she made a total of 150 recordings between 1904 and 1926, including a varied repertoire, but focusing mainly on opera \cite{Shawe-Taylor2001Melba}. One of Melba's most celebrated roles was Marguerite from Gounod’s \textit{Faust}. As a representative example, we selected Marguerite's aria “Ah! Je ris je me voir si belle”, a.k.a.~“Jewel aria”, from \textit{Faust}.

We additionally restored a version of the same aria from the soprano Adelina Patti (1843–1919), who sang the same repertoire as Melba, and their voices had similarities \cite{Forbes2001}. Still, we used a different voice, \textit{Female-5}, for fine-tuning because, in this recording, the impression of her voice is darker and lower compared to Melba's. 
The main reason for this difference might be that, at the time of the recording, Patti was already over 60 years old \cite{Forbes2001}.

Despite our reliance on the written literature, the main strategy for finding pairs between reference and original singers was analytic listening. With the limited bandwidth and the disturbing coloration of the original medium, some imagination was required when figuring out the original color of the voice.
Consequently, evaluating the fidelity of our results posed challenges, leading us to focus on a qualitative analysis.

\subsection{Qualitative Analysis}\label{sec:qualitativeanalysis}

Figure \ref{fig:spectrograms} shows spectrograms of two restoration examples:
a 1.5-second excerpt from Enrico Caruso's ``Vesti la Giubba'' (Victor 6001-A, 1910) in (a) and (b), and a same-length excerpt from Nellie Melba's ``Jewel aria'' (Victor 88066, 1910) in (c) and (d).
Both recordings show a bandwidth limit of approximately 4\,kHz.
In both cases, the original recording effective bandwidth barely exceeds 4\,kHz.
It can be observed that the limited bandwidth affects more critically the soprano voice than the tenor, as the former is at least an octave higher. Therefore, the restored voice of Caruso is recognisable and the hallucinated overtones create a warmer and fuller sound. In Melba's case, the result is more controversial. Even if the overtones are clearly audible, the voice sounds slightly breathy and lacks its ringing. 
The more overtones are generated, the more of the singer's personal color can be heard from the examples yielding a more natural-sounding result, although, in this case, part of the color has been hallucinated by the generative model. In Melba's case, the generated timbre is slightly different from what we can hear on the recordings.

Thanks to the extensive range of vocal techniques included by VocalSet, BABE-2 demonstrates a notable capacity for adapting to the distinct formant characteristics of various vowels, as illustrated in the green-colored box in Figs.~\ref{fig:spectrograms}(c, d). 
However, it is observed that subtle portamentos and singer legato lines are sometimes truncated, as shown in the cyan-colored box in Fig.~\ref{fig:spectrograms}(a, b).
This behavior is expected as soft (low-energy) singing has a much lower weight than louder passages in the reconstruction cost.  Also, further research is needed on how to better identify the vowel modification of the soprano voice in particular, i.e., the rounding of vowels at high pitches.

One limitation of the proposed method is that it is not designed to tackle nonlinear degradations, such as harmonic distortion.
Thus, one would expect that the model would fail when it encounters distortion artifacts. However, this does not seem to always be the case.
The red box in Fig.~\ref{fig:spectrograms}(a, b) highlights an example where Caruso's loud singing provoked a significant amount of harmonic distortion in the recording. 
In this example, BABE-2 suppresses the distortion and regenerates a clean high-frequency spectrum.
We attribute this behavior to the effect of the noise regularization implemented in our method and introduced in Sec.~\ref{sec:noise_reg}.
 
Due to scale limitations, BABE-2 does not have language-level control of the singing voice distribution. Since the frequency range of certain consonants, such as fricatives, is outside the range captured in the original recording, BABE-2 hallucinates consonants at statistically plausible locations, but these may not correspond to the lyrics. This is the case of the incorrectly placed fricative in the yellow box in Fig.~\ref{fig:spectrograms}(a, b).

The white boxes in Fig.~\ref{fig:spectrograms}(c, d) show some of the consequences of the source separation preprocessing. The bleeding from the instrumental track is present in the restored recording, especially in the low-frequency region.
The lack of robustness of the source separation model is a limitation that should be approached in future work.
Nevertheless, we also notice that a side-effect of BABE-2 is that it additionally suppresses some of the bleeding between the harmonics,  producing an even cleaner vocal separation. This behavior is expected, as these artifacts are not included in the training data.

\vspace{-2pt}
\section{Conclusions}\label{sec:conclusions}
\vspace{-2pt}

This study introduces BABE-2, enhancing the restoration of historical music recordings through a generative equalization strategy.
Objective evaluations demonstrate its superiority over the prior version, particularly in enhancing recordings of iconic vocalists Enrico Caruso and Nellie Melba. In addition to music enthusiasts, BABE-2 allows researchers to get a more accurate auditory picture of specific features of a recording or interpretation.
 Although certain limitations are identified, indicating areas for future research, this work represents a notable advancement in the domain of historical music restoration. BABE-2 demonstrates unprecedented potential to revitalize aged audio recordings, achieving levels of clarity and fidelity previously unattainable.

\section{Acknowledgments}
We acknowledge the computational resources provided by the Aalto Science-IT project. We thank Bryn Louise for proofreading parts of this paper.

\bibliographystyle{IEEEbib}
\bibliography{refs} 

\begin{thebibliography}{10}

\bibitem{godsill_digital_1998}
S.~J. Godsill and P.~J.~W. Rayner,
\newblock {\em Digital {Audio} {Restoration}---A Statistical Model Based Approach},
\newblock Springer, 1998.

\bibitem{kob2019interprete}
M.~Kob and T.~A. Weege,
\newblock ``How to interpret early recordings? {A}rtefacts and resonances in recording and reproduction of singing voices,''
\newblock {\em Computational Phonogram Archiving}, pp. 335--350, 2019.

\bibitem{Valimaki2008}
V.~Välimäki, S.~González, O.~Kimmelma, and J.~Parviainen,
\newblock ``Digital audio antiquing---{S}ignal processing methods for imitating the sound quality of historical recordings,''
\newblock {\em J. Audio Eng. Soc.}, vol. 56, no. 3, pp. 115--139, Mar. 2008.

\bibitem{stockham1975blind}
T.~G. Stockham, T.~M. Cannon, and R.~B. Ingebretsen,
\newblock ``Blind deconvolution through digital signal processing,''
\newblock {\em Proc. IEEE}, vol. 63, no. 4, pp. 678--692, 1975.

\bibitem{moliner2022two}
E.~Moliner and V.~V{\"a}lim{\"a}ki,
\newblock ``A two-stage {U}-net for high-fidelity denoising of historical recordings,''
\newblock in {\em Proc. IEEE Int. Conf. Acoust. Speech Signal Process. (ICASSP)}, Singapore, May 2022, pp. 841--845.

\bibitem{lemercier2024diffusion}
J.-M. Lemercier, J.~Richter, S.~Welker, E.~Moliner, V.~V{\"a}lim{\"a}ki, and T.~Gerkmann,
\newblock ``Diffusion models for audio restoration,''
\newblock {\em arXiv preprint arXiv:2402.09821}, 2024.

\bibitem{moliner2023zeroshot}
E.~Moliner, F.~Elvander, and V.~Välimäki,
\newblock ``Blind audio bandwidth extension: A diffusion-based zero-shot approach,''
\newblock {\em arXiv}, 2024.

\bibitem{song2020score}
Y.~Song, J.~Sohl-Dickstein, D.~P Kingma, et~al.,
\newblock ``Score-based generative modeling through stochastic differential equations,''
\newblock in {\em Proc. Int. Conf. Learning Representations (ICLR)}, May 2021.

\bibitem{karras2022elucidating}
T.~Karras, M.~Aittala, T.~Aila, and S.~Laine,
\newblock ``Elucidating the design space of diffusion-based generative models,''
\newblock {\em Adv. Neural Inf. Process. Syst. (NeurIPS)}, Dec. 2022.

\bibitem{karras2020analyzing}
T.~Karras, S.~Laine, M.~Aittala, et~al.,
\newblock ``Analyzing and improving the image quality of {StyleGAN},''
\newblock in {\em Proc. IEEE/CVF Conf. Computer Vision and Pattern Recognition}, Jun. 2020, pp. 8110--8119.

\bibitem{moliner2022solving}
E.~Moliner, J.~Lehtinen, and V.~V{\"a}lim{\"a}ki,
\newblock ``Solving audio inverse problems with a diffusion model,''
\newblock in {\em Proc. IEEE Int. Conf. Acoust. Speech Signal Process. (ICASSP)}, Rhodes, Greece, Jun. 2023.

\bibitem{chung2022diffusion}
H.~Chung, J.~Kim, M.~T. Mccann, et~al.,
\newblock ``Diffusion posterior sampling for general noisy inverse problems,''
\newblock in {\em Proc. Int. Conf. Learning Representations (ICLR)}, Kigali, Rwanda, May 2023.

\bibitem{chung2022parallel}
H.~Chung, J.~Kim, S.~Kim, and J.~C. Ye,
\newblock ``Parallel diffusion models of operator and image for blind inverse problems,''
\newblock in {\em Proc. IEEE Comput. Soc. Conf. Comput. Vis. Pattern Recognit. (CVPR)}, Vancouver, BC, Canada, Jun. 2023, pp. 6059--6069.

\bibitem{sadat2023cads}
S.~Sadat, J.~Buhmann, D.~Bradely, O.~Hilliges, and R.M Weber,
\newblock ``{CADS}: Unleashing the diversity of diffusion models through condition-annealed sampling,''
\newblock in {\em Proc. Int. Conf. Learning Representations (ICLR)}, 2024.

\bibitem{kingma_adam_2017}
D.~P. Kingma and J.~Ba,
\newblock ``Adam: {A} method for stochastic optimization,''
\newblock in {\em Proc. Int. Conf. Learn. Represent. (ICLR)}, San Diego, CA, May 2015.

\bibitem{wright2020perceptual}
A.~Wright and V.~V{\"a}lim{\"a}ki,
\newblock ``Perceptual loss function for neural modeling of audio systems,''
\newblock in {\em Proc. IEEE Int. Conf. Acoust. Speech Signal Process.}, 2020, pp. 251--255.

\bibitem{moliner2022diffusion}
E.~Moliner and V.~V{\"a}lim{\"a}ki,
\newblock ``Diffusion-based audio inpainting,''
\newblock {\em J. Audio Eng. Soc.}, vol. 72, Mar. 2024.

\bibitem{hawthorne2018enabling}
C.~Hawthorne, A.~Stasyuk, A.~Roberts, et~al.,
\newblock ``Enabling factorized piano music modeling and generation with the {MAESTRO} dataset,''
\newblock in {\em Proc. Int. Conf. Learning Representations (ICLR)}, May 2019.

\bibitem{kilgour2019frechet}
K.~Kilgour, M.~Zuluaga, D.~Roblek, and M.~Sharifi,
\newblock ``Fr{\'e}chet audio distance: A reference-free metric for evaluating music enhancement algorithms,''
\newblock in {\em Proc. Interspeech}, Aug. 2019, pp. 2350--2354.

\bibitem{gui2023adapting}
A.~Gui, H.~Gamper, S.~Braun, and D.~Emmanouilidou,
\newblock ``Adapting {F}rechet audio distance for generative music evaluation,''
\newblock in {\em Proc. IEEE Int. Conf. Acoust. Speech Signal Process. (ICASSP)}, 2024.

\bibitem{wu2023large}
Y.~Wu, K.~Chen, T.~Zhang, Y.~Hui, T.~Berg-Kirkpatrick, and S.~Dubnov,
\newblock ``Large-scale contrastive language-audio pretraining with feature fusion and keyword-to-caption augmentation,''
\newblock in {\em Proc. IEEE Int. Conf. Acoust. Speech Signal Process. (ICASSP)}, 2023.

\bibitem{defossez2022high}
A.~D{\'e}fossez, J.~Copet, G.~Synnaeve, and Y.~Adi,
\newblock ``High fidelity neural audio compression,''
\newblock {\em Transactions on Machine Learning Research}, 2023.

\bibitem{moliner2022behm}
E.~Moliner and V.~V{\"a}lim{\"a}ki,
\newblock ``{BEHM-GAN}: Bandwidth extension of historical music using generative adversarial networks,''
\newblock {\em IEEE/ACM Trans. Audio Speech Lang. Process.}, vol. 31, pp. 943--956, Jul. 2023.

\bibitem{wilkins2018vocalset}
J.~Wilkins, P.~Seetharaman, A.~Wahl, and B.~Pardo,
\newblock ``Vocalset: A singing voice dataset,''
\newblock in {\em ISMIR}, 2018, pp. 468--474.

\bibitem{rouard2023hybrid}
S.~Rouard, F.~Massa, and A.~D{\'e}fossez,
\newblock ``Hybrid transformers for music source separation,''
\newblock in {\em Proc. IEEE Int. Conf. Acoust. Speech Signal Process. (ICASSP)}, 2023.

\bibitem{Steane2002}
J.B Steane,
\newblock ``{Fach},''
\newblock {\em Groove Music Online}, 2002.

\bibitem{freestone1961enrico}
J.~Freestone,
\newblock {\em Enrico Caruso: His Recorded Legacy},
\newblock TS Denison, Minneapolis, 1961.

\bibitem{gentili2021birth}
B.~Gentili,
\newblock ``The birth of ‘modern’ vocalism: The paradigmatic case of {Enrico Caruso},''
\newblock {\em J. Royal Musical Assoc.}, vol. 146, no. 2, pp. 425--453, 2021.

\bibitem{Celletti2013}
R.~Celletti and A.~Blyth,
\newblock ``{Caruso, Enrico},''
\newblock {\em Groove Music Online}, 2013.

\bibitem{Shawe-Taylor2001Gigli}
D.~Shawe-Taylor and A.~Blyth,
\newblock ``{Gigli, Beniamino},''
\newblock {\em Groove Music Online}, 2001.

\bibitem{strong1941john}
L.~A.~G. Strong,
\newblock ``{John McCormack: The story of a singer},''
\newblock {\em The Macmillan Company}, 1941.

\bibitem{Shawe-Taylor2001Melba}
D.~Shawe-Taylor,
\newblock ``{Melba, Dame Nellie},''
\newblock {\em Groove Music Online}, May 2009.

\bibitem{Forbes2001}
E.~Forbes,
\newblock ``{Patti, Adelina},''
\newblock {\em Groove Music Online}, 2001.

\bibitem{huang2021multi}
R.~Huang, F.~Chen, Y.~Ren, J.~Liu, C.~Cui, and Z.~Zhao,
\newblock ``Multi-singer: Fast multi-singer singing voice vocoder with a large-scale corpus,''
\newblock in {\em Proceedings of the 29th ACM International Conference on Multimedia}, 2021, pp. 3945--3954.

\bibitem{wang2022opencpop}
Y.~Wang, X.~Wang, P.~Zhu, J.~Wu, H.~Li, H.~Xue, Y.~Zhang, L.~Xie, and M.~Bi,
\newblock ``Opencpop: A high-quality open source chinese popular song corpus for singing voice synthesis,''
\newblock in {\em Interspeech}, 2022.

\bibitem{duan2013nus}
Z.~Duan, H.~Fang, B.~Li, K.~C. Sim, and Y.~Wang,
\newblock ``The {NUS} sung and spoken lyrics corpus: A quantitative comparison of singing and speech,''
\newblock in {\em Proc. Asia-Pacific Signal and Information Processing Association Annual Summit and Conference}, 2013, pp. 1--9.

\bibitem{koguchi2020pjs}
J.~Koguchi, S.~Takamichi, and M.~Morise,
\newblock ``{PJS}: Phoneme-balanced japanese singing-voice corpus,''
\newblock in {\em Proc. Asia-Pacific Signal and Information Processing Association Annual Summit and Conference}, 2020, pp. 487--491.

\bibitem{zhang2022m4singer}
L.~Zhang, R.~Li, S.~Wang, L.~Deng, J.~Liu, Y.~Ren, J.~He, R.~Huang, J.~Zhu, X.~Chen, et~al.,
\newblock ``M4singer: A multi-style, multi-singer and musical score provided mandarin singing corpus,''
\newblock {\em Adv. Neural Inf. Process. Syst. (NeurIPS) Datasets and Benchmarks Track}, vol. 35, pp. 6914--6926, 2022.

\bibitem{choi2020children}
S.~Choi, W.~Kim, S.~Park, S.~Yong, and J.~Nam,
\newblock ``Children’s song dataset for singing voice research,''
\newblock in {\em Proc. ISMIR}, 2020, vol.~4.

\bibitem{yang2024guidance}
Lingxiao Yang, Shutong Ding, Yifan Cai, Jingyi Yu, Jingya Wang, and Ye~Shi,
\newblock ``Guidance with spherical gaussian constraint for conditional diffusion,''
\newblock {\em arXiv preprint arXiv:2402.03201}, 2024.

\bibitem{ho2022video}
J.~Ho, T.~Salimans, A.~Gritsenko, W.~Chan, M.~Norouzi, and D.~J. Fleet,
\newblock ``Video diffusion models,''
\newblock {\em Adv. Neural Inf. Process. Syst. (NeurIPS)}, Dec. 2022.

\end{thebibliography}

\appendix

\section{Supplementary Algorithm specifications} \label{sec:appendix_details}

This appendix includes additional specifications and implementation insights regarding the BABE-2 algorithm. Many aspects covered here were omitted from the main paper either because they build upon established work or due to their detailed, technical nature.

\subsection{Training} \label{sec:setup}

The training objective is based on the one expressed in Eq.~\ref{loss}.
However, we apply the preconditioning strategy proposed by Karras et al. \cite{karras2022elucidating}, which aimed to improve the diffusion model training dynamics.
The denoiser in Eq. \ref{loss} is reparameterized as:
\begin{equation}\label{eq:precondition}
D_\theta(\bm{x}_\tau, \tau)=
c_\text{skip}(\tau)\bm{x}_\tau+
c_\text{out}(\tau)F_\theta(c_\text{in}(\tau)\bm{x}_\tau, \tfrac{\ln(\sigma(\tau))}{4}),
\end{equation}
where $c_\text{in}$,  $c_\text{skip}$ and $c_\text{out}$ are time-dependent preconditioning parameters defined in \cite{karras2022elucidating}, and $F_\theta$ corresponds to the optimizable neural network architecture, as detailed in Sec. \ref{sec:architecture}.
The loss weighting term $\lambda(\tau)$ in Eq. \ref{loss} is defined as: $\lambda(\tau)=1/c_\text{out}(\tau)^2$, also following the guidelines from \cite{karras2022elucidating}.

The diffusion model employed for piano restoration was trained using the MAESTRO dataset \cite{hawthorne2018enabling}.
For the singing voice experiments, we pretrained a new diffusion model with a collection of studio quality modern singing voice recordings from different sources \cite{huang2021multi, wang2022opencpop,  duan2013nus, koguchi2020pjs, zhang2022m4singer, choi2020children}.
This pre-training phase spanned 325k iterations over three days on a single NVIDIA A100-80GB GPU.
The singing voice models were later fine-tuned with a smaller reference set extracted from selected singers from Vocalset \cite{wilkins2018vocalset},
as we elaborate on in Sec.~\ref{sec:voice_analysis}.
We employed standard fine-tuning, resuming the training from the pre-trained weights. The fine-tuning process involved 8k training iterations, requiring about two hours on the same GPU.

The preconditioning parameters used in Eq. \ref{eq:precondition}
depend on the average power of the training data $\sigma_\text{data}^2$.
For the MAESTRO dataset, this value was estimated as $\sigma_\text{data}=6.3\times10^{-2}$.
For the singing voice experiments in particular, as we were working with heterogeneous datasets, we decided to normalize the training data samples.
During the pre-training stage, we apply Root-Mean-Square (RMS) normalization, thus resulting with the parameter $\sigma_\text{data}=1$. 
For fine-tuning, in order to preserve the volume-dependent characteristics of singing voice, we estimated the average RMS value from a representative set of the fine-tuning dataset, in our case VocalSet, and normalize all the training samples with it. For this purpose, we estimated the value $\sigma_\text{norm}=0.04$.

Additionally, as often employed for training diffusion models, we track an exponential moving average (EMA) of the neural network weights during training. The EMA weights are later used during the inference stage. Some of the most important parameters regarding training are summarized in Table\,\ref{tab:training}.



\begin{table}[]
\caption{Training hyperparameters chosen in our experiments}
\label{tab:training}
\begin{tabular}{@{}llll@{}}
\toprule
\textbf{Training params.}    & Piano & S. Voice (pre) & S. Voice (f-t)  \\ \midrule
Learning rate & $2\times 10^{-4}$ & 2$\times 10^{-5}$ & $1\times 10^{-4}$ \\
Batch size  & 4 & 4 & 4 \\
Sampling rate &22.05 kHz & 44.1 kHz & 44.1kHz \\
Segment length &8.35 s& 5.94 s & 5.94 s \\
EMA rate &0.9999& 0.9999& 0.9999\\
$\sigma_\text{data}$ & $6.3 \times 10^{-2}$ &1 &1  \\
Training its. & 850k & 325k &  8k\\ 
\bottomrule

\end{tabular}%
\end{table}

\subsection{Neural Network Architecture}\label{sec:architecture}

In this study, we employed the CQT-Diff+ architecture \cite{moliner2022diffusion} as the backbone for our diffusion model, denoted as $F_\theta$.
This architecture incorporates optimizable layers with an invertible Constant-Q Transform (CQT), 
 characterized by varying time resolutions across octave bands. 
The architecture adopts a U-Net-style encoder-decoder structure with skip connections, primarily consisting of 2D-Convolutional layers that process along both time and frequency axes.

For our experiments involving piano music and singing voice, we utilized distinct parameter configurations, yet both setups comprise approximately 40 million trainable parameters. 
Specifically, the piano music configuration aligns with that employed in BABE \cite{moliner2023zeroshot}, utilizing a CQT that spans 7 octaves with 64 frequency bins per octave. 
 In contrast, the singing voice experiments required adaptation to a higher sampling rate, resulting in a CQT covering 8 octaves but with 32 frequency bins per octave. 
 This adjustment in frequency resolution, particularly for singing voice, proved beneficial for generating a more coherent spectrum in the produced audio, a critical aspect for tasks like equalization.
 Furthermore, an important consideration in both experimental configurations is the exclusion of frequencies below 86 Hz, achieved through high-pass filtering.
 While this decision prevents the model from restoring content within this low-frequency range, it is not deemed problematic for our purposes. This is because the musical signals we aim to process typically do not contain significant information in such low frequencies.

\subsection{Inference}

The inference algorithm was designed to share the same structure as BABE \cite{moliner2023zeroshot}, but including BABE-2's new contributions detailed in the main paper. Some of the most relevant hyperparameters employed for inference are summarized in Table \ref{table:hyperparameters}.

We use the exponentially-warped discretization from \cite{karras2022elucidating}:
\begin{equation}\label{schedule}
    \tau_{i<T}=\left(\sigma_{\text{start}}^{\;\;\;\frac{1}{\rho}} 
    + \tfrac{i}{T-1}\left(
    \sigma_\text{min}^{\;\;\;\frac{1}{\rho}}
    -\sigma_\text{start}^{\;\;\;\frac{1}{\rho}}
    \right)\right)^\rho,
\end{equation}
where the hyperparameters $\sigma_\text{start}$, $\sigma_\text{min}$, $T$ and $\rho$ define the time schedule. Intuitively, the starting noise level $\sigma_\text{start}$ should be large enough to mask all the imperfections from the audio signal used at initialization, and $\sigma_\text{min}$ should be negligible when compared with the audio signal magnitude. 
The role of the parameter $\rho$ is to warp the time discretization and concentrate more steps towards lower noise levels. 
The parameter $T$ defines the number of discretized steps. With it, one can effectively trade-off audio quality per inference speed. While using a larger $T$ can potentially lead to better results, we decied to fix $T=51$ throughout all our experiments.
It must be noted that, since we employ a second-order sampler, the number of neural function evaluations required during inference is twice the value of $T$.



Following \cite{moliner2022solving, moliner2023zeroshot}, we normalize the likelihood score step size $\xi(\tau)$ as follows to enhance robustness:
 \begin{equation}
\xi(\tau)=\xi^\prime \sqrt{N}/( \sigma(\tau) \lVert \nabla_{\bm{x}_\tau} \lVert \bm{y} - \textbf{m}\odot\bm{\hat{x}}_0 \rVert^2 \rVert^2),
 \end{equation}
a technique also recognized as a spherical Gaussian constraint in recent literature \cite{yang2024guidance}. 
The parameter $\xi^\prime$ plays a crucial role by balancing audio quality and fidelity to observations. Higher $\xi^\prime$ values aim to minimize the cost function effectively, although excessively large values may lead to artifacts in the audio.
We observed that introducing stochasticity in the inference phase allows enables the application of larger $\xi^\prime$ values while maintaining audio quality.
This stochasticity is introduced through the $S_\text{churn}$ parameter of the sampler, and the noise regularization from Sec.\,\ref{sec:noise_reg}.

The filter parameters are optimized through stochastic gradient descent using the \emph{Adam} optimizer. The learning rate was searched qualitatively by trial and error, and was defined to a value of 10.  We adopted the default momentum parameters from the \emph{PyTorch}'s implementation.
Optimizing the filter parameters is computationally less demanding than updating audio signals, as it avoids the need for neural functions evaluations.
Consequently, we employed a relatively high number of filter optimization iterations for each sampling step, specifically 100.


\begin{table}[]
\caption{Inference hyperparameters chosen in our experiments}
\label{table:hyperparameters}
\begin{tabular}{@{}lll@{}}
\toprule
\textbf{Inference params.}    & Piano & Singing Voice  \\ \midrule
Start noise level $\sigma_\text{start}$ & 0.5 & 10 \\
Min. noise level $\sigma_\text{min}$ & 4$\times 10^{-5}$ & 1$\times 10^{-3}$ \\
Time warping $\rho$ & 13 & $13$  \\
Discretization steps $T$ & 51 & $51$  \\
Sampling stochasticity $S_\text{churn}$ & 10 & $10$  \\
Step size $\xi^\prime$ &                 1.0 &    0.5                            \\
Noise regularization $\gamma$ & 0.25 & $1$  \\
Filter learning rate  &      10           & 10                                \\ 
Num. optimization its. per step  &      100           & 100                                \\ 
BCR decay rate $\beta$ & $0.1$ & $0.1$  \\
BCR weight  $\gamma_\text{BCR}$ & $10$ & $10$  \\
\bottomrule

\end{tabular}%
\end{table}

\subsection{Filter Parameterization}

The proposed filter parameterization is introduces in Sec. \ref{sec:filter_parameterization}. The piece-wise linear transfer function can be formalized as follows:
\begin{equation}\label{filter}
H_\phi(f) [\text{dB}] = \begin{cases}
  A_\text{lim-} \log_2 \frac{f}{f_{\text{-}S^\prime}} + \sum_{i=\text{-}S^\prime}^{\text{-}1} A_i &   f <  f_{\text{-}S^\prime} \\
  \hspace{30pt}\vdots  & \hspace{25pt}\vdots \\
  A_{\text{-}2} \log_2 \frac{f}{f_{\text{-}1}} +A_\text{-1} +A_\text{-2}&  f_{\text{-}2} \leq f < f_{\text{-}1}  \\
  A_{\text{-}1} \log_2 \frac{f}{f_{0}} &  f_{\text{-}1} \leq f < f_{0}  \\
  A_1 \log_2 \frac{f}{f_{0}} &  f_{0} \leq f < f_{1}  \\
  A_2 \log_2 \frac{f}{f_{1}} +A_1 &  f_{1}  \leq f < f_{2} \\
  \hspace{30pt}\vdots  & \hspace{25pt}\vdots \\
  A_\text{lim+} \log_2 \frac{f}{f_{S}} + \sum_{i=1}^{S} A_i &  f_{S}  \leq f,  \\
\end{cases}
\end{equation}
where $f_{i}$ (Hz) represent cutoff frequencies and $A_i$ (dB) are the decay slopes. The set of adjustable parameters is defined as: 
\begin{equation}
   \phi=\{ f_{i} , A_i \mid i=1,\dots, S \},
\end{equation}
where $S$ is the number of breakpoints.

For the reasons specified in Sec. \ref{sec:filter_parameterization}, we use a reduced number of $S=5$ stages in our experiments. This comprises a total of 9 optimizable parameters. We initialized the 5 breakpoints to, and the rest to: $f_{-2}$\,=\,50\,Hz, $f_{-1}$\,=\,500\,Hz, $f_0$\,=\,1\,kHz, $f_1$\,=\,1.5\,kHz, and $f_2$\,=\,2\,kHz, while all the slopes are initialized to 0\,dB.
Considering the the $A_\text{lim+}$ slope is fixed at -80\,dB, this initialization produces a flat magnitude response with a steep band-limit at 2\,kHz, a relatively low cut-off frequency.
The initialization plays a relevant role in the performance of the method. However we notice that, in this case, the filter initialization influence is not as critical as in the prior version, BABE. We attribute this observed improvement in robustness to some of the incorporated additions, specifically the noise regularization and the Adam optimizer.


\subsection{Block-Autoregressive Inference}\label{sec:BAI}




Due to memory constraints, the diffusion models employed in this work are designed to process short audio segments within the range of a few seconds. However, in the context of audio restoration, it is necessary to be able to restore entire recordings which may last several minutes long.

Adapting diffusion models to function on a frame-by-frame basis while ensuring coherence between subsequent frames is a known trick in the field \cite{ho2022video}. 
One can just enforce consistency with the last fragment of the previous frame by incorporating an ``outpainting'' objective on top of the reconstruction.
Given the last fragment of the previous restored frame repositioned at the start of the current one $\mathbf{x}_\text{prev}$, the proposed algorithm can be adapted to a block-autoregressive inference scenario by substituting the reconstruction cost function in Eq. \eqref{eq:rec} for:
\begin{multline}
    C_\text{audio}^\text{B-AR}(\mathbf{y}, 
    \mathbf{x}_\text{prev}, 
    H_{\phi}(\hat{\mathbf{x}}_0))= 
    \lVert
   \mathbbm{1}_{[0,t_\text{ov.}]} \odot(\mathbf{x}_\text{prev}-
   \hat{\mathbf{x}}_0) \\
+  \mathbbm{1}_{[t_\text{ov.},t_\text{end}]}\odot(\mathbf{y} -
    H_\phi(\mathbf{\hat{x}}_0)
    )
    )
    \rVert_2^2,
\end{multline}
where $\mathbbm{1}_{[t_\text{start}, t_\text{end}]}$ represents a binary mask that is equal to 1 in the time interval between $t_\text{start}$ and $t_\text{end}$ and 0 otherwise, $t_\text{ov.}$ is the overlap time, $t_\text{end}$ points to the end of the segment, and $\odot$ denotes element-wise multiplication. 
In our experiments we use an overlap factor $t_\text{ov.}/t_\text{end}$ of 10\%.
 This trick was already employed in BABE \cite{moliner2023zeroshot}, but, although mentioned, the adapted cost function for bandwidth extension was not properly formalized, and neither is anywhere else in the related literature, as far as we are aware.

This adaptation can be equivalently applied to the filter reconstruction cost function in Eq. \eqref{eq:filter_rec}, allowing us to estimate the filter parameters with the updated objective.
Note however that, if the degradation is considered to be time-invariant, there is no need to re-estimate the filter parameters $\phi$ among different frames, as only once is needed.
Nevertheless, we use the block-autoregressive inference in our experiments for evaluation purposes.
\end{document}